\documentclass[conference]{IEEEtran}
\IEEEoverridecommandlockouts
\usepackage{cite}
\usepackage{amsmath,amssymb,amsfonts}
\usepackage{algorithmic}
\usepackage{graphicx}
\usepackage{textcomp}
\usepackage{xcolor}
\def\BibTeX{{\rm B\kern-.05em{\sc i\kern-.025em b}\kern-.08em
    T\kern-.1667em\lower.7ex\hbox{E}\kern-.125emX}}
    
\usepackage{scalerel}
\usepackage{tikz}
\usetikzlibrary{svg.path}
\definecolor{orcidlogocol}{HTML}{A6CE39}
\tikzset{
  orcidlogo/.pic={
    \fill[orcidlogocol] svg{M256,128c0,70.7-57.3,128-128,128C57.3,256,0,198.7,0,128C0,57.3,57.3,0,128,0C198.7,0,256,57.3,256,128z};
    \fill[white] svg{M86.3,186.2H70.9V79.1h15.4v48.4V186.2z}
                 svg{M108.9,79.1h41.6c39.6,0,57,28.3,57,53.6c0,27.5-21.5,53.6-56.8,53.6h-41.8V79.1z M124.3,172.4h24.5c34.9,0,42.9-26.5,42.9-39.7c0-21.5-13.7-39.7-43.7-39.7h-23.7V172.4z}
                 svg{M88.7,56.8c0,5.5-4.5,10.1-10.1,10.1c-5.6,0-10.1-4.6-10.1-10.1c0-5.6,4.5-10.1,10.1-10.1C84.2,46.7,88.7,51.3,88.7,56.8z};
  }
}
\newcommand\orcidicon[1]{\href{https://orcid.org/#1}{\mbox{\scalerel*{
\begin{tikzpicture}[yscale=-1,transform shape]
\pic{orcidlogo};
\end{tikzpicture}
}{|}}}}
\usepackage[hidelinks]{hyperref}    

\hypersetup{
pdftitle={Development of a Meta-language and its Qualifiable Implementation for the Use in Safety-critical Software},
pdfsubject={Doctoral Symposium - MODELS CONFERENCE 2021},
pdfauthor={Vanessa Tietz},
pdfkeywords={Ada SPARK, domain-specific modeling, model-based systems, engineering, qualifiable runtime environment, safety-critical}
}

\newcommand{\includevisio}[2][]{\includegraphics[clip, trim=0.09cm 0.2cm 0.3cm 0.3cm, #1]{#2}} 
    
\begin{document}

\title{Development of a Meta-language and its Qualifiable Implementation for the Use in Safety-critical Software}

\author{\IEEEauthorblockN{Vanessa Tietz \orcidicon{0000-0002-5942-5893}}
\IEEEauthorblockA{\textit{Institute of Aircraft Systems} \\
\textit{University of Stuttgart}\\
Stuttgart, Germany \\
0000-0002-5942-5893}
}

\maketitle
\makeatletter
\newcommand\copyrighttext{%
  \footnotesize \textcopyright 2021 IEEE. Personal use of this material is permitted.
  Permission from IEEE must be obtained for all other uses, in any current or future
  media, including reprinting/republishing this material for advertising or promotional
  purposes, creating new collective works, for resale or redistribution to servers or
  lists, or reuse of any copyrighted component of this work in other works.
  }
\newcommand\copyrightnotice{%
\begin{tikzpicture}[remember picture,overlay]
\node[anchor=south,yshift=10pt] at (current page.south) {\fbox{\parbox{\dimexpr\textwidth-\fboxsep-\fboxrule\relax}{\copyrighttext}}};
\end{tikzpicture}%
}
\copyrightnotice

\begin{abstract}
The use of domain-specific modeling for development of complex (cyber-physical) systems is gaining increasing acceptance in the industrial environment. Domain-specific modeling allows complex systems and data to be abstracted for a more efficient system design, development, validation, and configuration. However, no existing (meta-)modeling framework can be used with reasonable effort in certified software so far, neither for the development of systems nor for the execution of system functions.
For the use of (development) artifacts from domain-specific modeling in safety-critical processes or systems it is required to ensure their correctness by either subsequent (manual) verification or the usage of (pre-)qualified software. Existing meta-languages often contain modeling elements that are difficult or impossible to implement in a qualifiable manner leading to a high manual, subsequent certification effort. 
Therefore, the aim is to develop a (meta-)modeling framework, that can be used in certified software. This can significantly reduce the development effort for safety-critical systems and enables the full advantages of domain-specific modeling. The framework components considered in this PhD-Thesis include: (1) an essential meta-language, (2) a qualifiable runtime environment, and (3) a suitable persistence. The essential \mbox{(meta-)}modeling language is mainly based on the UML standard, but is enhanced with multi-level modeling concepts such as deep instantiation. Supporting a possible qualification, the meta-language is implemented using the highly restrictive, but formally provable programming language Ada SPARK.

\end{abstract}

\section{Problem and Motivation}
Domain-specific modeling (DSM) is commonly used for the development of complex cyber-physical systems and software. In DSM, the model elements represent objects of the real application domain. This allows complex systems and data to be abstracted in order to achieve a more efficient design, development, validation and configuration. The higher abstraction level of domain-specific models allows complex structures and relationships to be represented and efficient automation processes like the automatic generation of development artifacts to be introduced. (Meta-)modeling frameworks can be used in the development of (cyber-physical) systems as well as in software for the execution of system functions handling complex data structures. A variety of frameworks, such as the Eclipse Modeling Framework (EMF) \cite{Steinberg2009}, the Generic Modeling Environment (GME) \cite{GME}, Matlab Simulink \cite{MATLAB_SIMULINK} and System Composer \cite{MATLAB_SC} or SCADE \cite{SCADE} are available for DSM. 
Nevertheless, there are only a few examples where the artifacts created with DSM are utilized to support the certification process or as part of safety-critical software. This is mainly due to the fact that the required subsequent (manual) qualification effort to ensure the correctness of the artifacts is disproportionately high for the user of the (meta-)modeling framework. For qualification, the user must provide proof that the (development) objectives specified in standards such as DO-178C \cite{DO178-C} or ISO26262 \cite{ISO26262} are complied with. This includes, proof of the absence of undesirable behavior, compliance with predefined requirements and proof of test coverage by means of coverage metrics. It is possible to reduce the subsequent qualification effort by using a pre-qualified (meta-)modeling framework or at least components of the framework assisting the tool user in conducting the qualification processes. Therefore, the aim of the PhD-Thesis is to develop parts of a qualifiable (meta-)modeling framework applicable in safety-critical software. In this context, the term framework is related to a tool that can be used in (safety-critical) software, which at least consists of a meta-language and a runtime. The (meta-)modeling framework shall support the development process of safety-critical systems as well as system functions where complex data models are needed e.g. predictive maintenance or artificial intelligence. 
All implied problems of current (meta-)modeling frameworks are described in more detail below to motivate research questions. \\

\textit{\textbf{Problem 1} - Many meta-modeling languages are hardly usable in the safety-critical domain:}
Existing meta-languages are general-purpose and, therefore highly complex in their structure, and contain modeling elements whose implementation might not be qualified due to imprecisely defined or non-deterministic behavior in the implementation. If the artifacts created with such a meta-language can be qualified, then only with high subsequent (manual) effort. Elements that are difficult to qualify include, for example, opposite-references and proxies, which are very complex in their behavior and thus often cannot be programmed deterministically. 
The selective reduction of the functional scope of existing languages via profiles and the subsequent generation of safety-critical code via a code generator is not sufficient. This is because the environments does without qualification artifacts such as requirements and design documents. This means that a new runtime environment has to be programmed anyway, documented according to the requirements derived from standards. Furthermore, the code generator would also have to be qualified accordingly. 
\fbox{\parbox{8.7cm}{\textbf{RQ1:} What does a meta-language look like whose implementation leads to no ambiguity or non-deterministic behavior at runtime?}}
\vspace{0.01cm}

\textit{\textbf{Problem 2} - Many runtime environments of meta-languages are implemented with object-oriented programming languages or in an object-oriented manner:} 
Software implemented with an object-oriented programming language can barely be qualified due to issues like dynamic binding, overloading, try and catch constructs, multiple inheritance or garbage collectors \cite{OOP_Issues}\cite{DO332}.
More issues regarding the application of object-oriented programming are listed in \cite{OOP_Issues} and \cite{DO332}. Additionally, programming languages (including procedural ones) such as C++, Java or C do not provide an internal mechanism for formal proof of semantic correctness that could reduce the (manual) qualification effort.
\fbox{\parbox{8.7cm}{\textbf{RQ2:} How and with which programming language can a formally verifiable runtime environment for meta-models and domain-specific models be implemented? What does such an implementation look like and what has to be done to reduce the manual qualification effort?}}
\vspace{0.01cm}

\textit{\textbf{Problem 3} - Reflective requests can only be qualified under huge manual effort so far and therefore are hardly used in safety-critical applications:}
A powerful feature of DSM is that meta-models can be read and modified during run-time. It is the specific intention to enable this feature via reflective requests also for safety-critical systems in order to achieve greater flexibility in possible areas of application. Reflection means that a program can know and possibly modify its own structure. In this case, the reflections refer to the created domain-specific models whose structure is the meta-model. There is a risk of manipulating data in such a way that it is no longer consistent with the underlying model. In addition, all meta-interactions have to be deterministic and an unintended loss of data due to reflective requests must be avoided at all expense.  
\fbox{\parbox{8.7cm}{\textbf{RQ3:} How can we ensure that changes to the meta-model at runtime via reflective requests do not lead to undesired misbehavior and corrupted data?}}

\section{Related Work}
A variety of frameworks are available for DSM such as the Eclipse Modeling Framework (EMF) \cite{Steinberg2009}, the Generic Modeling Environment (GME) \cite{GME} or Matlab Simulink \cite{MATLAB_SIMULINK} and System Composer \cite{MATLAB_SC}. These frameworks allow the efficient mapping of complex relationships in application-specific data models using a meta-language (e.g..: UML \cite{UML}, or (E)MOF \cite{MOF}), a meta-model (e.g.: OAAM \cite{OAAM}, SysML \cite{SysML} or AADL \cite{AADL}) and a domain-specific model. Due to the high abstraction level of domain-specific models, development artifacts e.g. source code can be created. Therefore, domain-specific models are predestined for application in the development of complex, cyber-physical systems. 
For the use of such frameworks and languages in the safety-critical domain, the output must be verified (manually), using model checkers, qualified (validation) software or manual review activities accordingly to standardized processes or deliverables. In the field of avionics, for example, such deliverables are described in DO-178C \cite{DO178-C}, DO-330 \cite{DO330}, DO-331 \cite{DO331}, DO-332 \cite{DO332}, and DO-333 \cite{DO333}, or for the automotive sector in ISO 26262 \cite{ISO26262}. The processes include writing requirements, choosing a programming language with a qualified compiler, complying with coding standards and guidelines like MISRA-C \cite{MISRAC}, proving the absence of undesired behavior and the fulfillment of requirements, proving test coverage by means of dedicated coverage metrics (e.g. modified-decision coverage), and performing review processes \cite{DO178-C}. Due to the high effort resulting from the mandatory qualification processes, the utilization of domain-specific models or frameworks in qualified software or development processes is comparatively rare with respect to their use in the non-critical cyber-physical domain. The latter statement targets the use of DSMs in the qualified path not as optional auxiliary tools. 
The most common approaches in the safety-critical domain are attempts to use domain-specific models to automatically generate safety-critical software, which is outside the scope under consideration. This includes, for example, SCADE \cite{SCADE} or TargetLink \cite{TargetLink}. Furthermore, there are many approaches to facilitate the use of models in the safety-critical area via subsequent formal verification and validation \cite{Van_Acker_2020} \cite{VanAcker2018} of models and requirements through model-checking \cite{Zalila2013} \cite{Haxthausen2012} \cite{zalila2014methods}. In addition to a model checker, the UPPAAL tool also offers the possibility of creating abstract models, simulate dynamic behavior, and specify and verify the safety of the models \cite{UPPAAL} \cite{larsen1997uppaal}. Another approach is OSATE \cite{OSATE}, which supports the validation of AADL \cite{AADL} models according to all naming and legality rules defined in the AADL \cite{AADL} standard. Additionally, it provides code templates, real-time syntax checking, code completion, and proposals to fix errors. Ocarina \cite{Lasnier2009}, a stand-alone model processor written in the Ada programming language, is also based on AADL. The Ocarina tool suite provides model manipulation, syntactic/ semantic analysis, and the verification and code generation from AADL models. Model verification without direct reference to safety-critical systems is addressed in \cite{Besnard_2020} \cite{Besnard2018} \cite{Hilken2014}.
I am unaware of any (meta)-modeling framework directly usable in qualified software.
Almost all mentioned approaches are based on existing (meta-)modeling frameworks and languages and achieve applicability in safety-critical software by e.g. simplifying qualification processes through model verification and validation or subsequent qualified transformations (automatic code generation). None of them considers qualifiability by introducing a new qualifiable implementation of a meta-language. With our approach the outcome of modeling shall be directly usable without the need of subsequent qualification processes. This enables the utilization of (meta-)modeling in real-time safety-critical system functions.

\section{Proposed Solution}
To solve the related issues and to answer the research questions, a new (meta-)modeling framework applicable in safety-critical software will be developed. The framework shall consist of (1) an essential meta-language, (2) a qualifiable runtime, (3) a suitable persistence, (4) a deterministic transformation language, (5) an automatic generation of qualification artifacts, and (6) a decoupled visualization. In order to facilitate a possible qualification, all mentioned components should work decoupled and can only communicate with each other via a generic interface. This allows the individual components to be viewed and (pre-)qualified independently if it is desired. At the same time, this means that individual components within the framework should be replaceable by other implementations. In this PhD-Thesis the components (1) to (3) are considered and described in detail. Since there are always different notations in the field of meta-modeling, the following notation will be used: The meta-language corresponds to the M3-level, the meta-model to the M2-level and the domain-specific model to the M1-level.\\
\textbf{(1) Essential Meta-Language:} The most important aspect about the meta-language is the avoidance of modeling elements that cannot be implemented deterministically in the runtime environment. This addresses \textbf{RQ1}. In addition, the focus is to simplify the complexity of the language by only providing essential functionalities that are needed for static modeling of complex data and system structures. This means, for example, that there is no need to define operations nor to enable software modeling. The goal is to create alternatives for elements that are difficult to qualify where necessary. To obtain a correct understanding of this language and the models generated on M2 and M1 level, an unambiguous interpretability of the language as well as of the M2 and M1 models will be taken into account. Moreover, the language should provide both an interface for the automatic generation of qualification artifacts and the ability to define constraints at M2 level which can be proven on M1 level.\\  
\textbf{(2) Qualifiable Runtime:} The runtime environment, addressing \textbf{RQ2}, will provide the functionality of creating, editing and, in general, interacting with meta-models on M2 level and domain-specific models on M1 level based on the semantic and rules defined by the meta-language on M3 level. The runtime is exclusively a model editor and not a model executor. Similar to the meta-language, minimalism and simplicity is emphasized. In this case regarding the number of code lines - the larger the amount of code, the more has to be checked for a possible qualification. Additionally, no automatic generation of code from created meta-models is provided, the runtime environment operates only as a model interpreter which is implemented with a qualifiable programming language. Another possibility to reduce the qualification effort to a minimum is the choice of a suitable, formally provable and restrictive programming language. 
Even if the chosen programming language supports object-oriented programming, the management of objects during runtime should be conducted via an object-relational mapping approach. This initially leads to a greater computational effort, but is easier to qualify because this type of object management is based on easily comprehensible operations based on list entries and links to each other via global unique identifiers. The interaction with the models on M2 and M1 level shall be based on atomic CRUD (Create, Read, Update, Delete) requests via a sole and unambiguous interface to the other components. When implementing the meta-language, care should be taken to ensure that the integrity of the data is maintained at all times. \textbf{RQ3} is addressed during the implementation of the qualified runtime, too. The implementation of the needed reflective requests shall ensure data integrity and consistency of all models during their execution. \\
\textbf{(3) Suitable Persistence:} In order to store the created meta and domain-specific models, suitable deterministic and reliable serialization processes should be implemented based on the type of data-storage used.

\section{Evaluation and Validation}
The functionality and operation of the meta-modeling framework will be demonstrated with a demonstrator. This demonstrator will be a software tool usable with a graphical interface. First of all, the correct implementation of the meta-language as well as the compliance with predefined requirements will be verified through the execution of several (pre-defined) use-cases. If the implementation is correct and all requirements have been met, the focus will be on the qualifiability of the runtime environment. A possible qualification process can be supported, via the proof of freedom from runtime errors by means of formal test methods and the proof of deterministic program behavior via program tracing. 
From a functional point of view, the usability of the modeling language, modeling possibilities, the runtime and process workload, qualification effort, as well as the clarity and structure of the models created will be considered and compared to state-of-the-art solutions like EMF or (Web)GME. For example, different use cases will be defined and modeled, and existing meta-models such as the OAAM or AADL will be recreated. 
During the complete development process, different versions of the demonstrator will be given to (industrial) partners in order to get feedback on the quality and usability of the framework in their special use-cases and to be able to respond to their requirements in an early stage. It is conceivable to define evaluation metrics by which the quality of the new framework is assessed by the (industrial) partners in comparison to existing frameworks and an evaluation is quantitatively feasible.

\section{Expected Contributions}
In this section a list of expected theoretical and practical contributions of the PhD research is provided.
\begin{itemize}
    \item The most fundamental contribution is the enabling of the development of a new (meta-)modeling framework that can be used in qualified software, making DSM more accessible for the development of safety-critical systems as well as for safety-critical system functions in e.g. avionics, or autonomous driving. 
    \item Comprehensive understanding of which elements of meta-languages make qualification cumbersome. This includes implementation options and their impact on the qualification process.
    \item An essential meta-language for DSM that is free from modeling elements that could lead to non-deterministic behavior in the implementation.  
    \item The implementation of the meta-language with a formal provable but strongly restrictive programming language. 
    \item Providing the option to use reflective requests in safety-critical software. 
    \item A deterministic and comprehensive definition of meta-model modifications during run-time.
    \item The (meta-)modeling framework usable in qualified software enables the usage in advanced system functions depending on complex data- and system models in the safety-critical domain e.g. artificial intelligence, predictive maintenance, plug $\&$ fly avionics, and autonomous driving.
\end{itemize}

\section{Current Status and Timeline}
To answer all defined research questions (\textbf{RQ1} - \textbf{RQ3}) different tasks have to be conducted. These are sorted and listed according to their component of the framework:\\
\textbf{Whole Framework:} (1) Defining requirements for a qualifiable DSM framework, (2) conception of a qualifiable (meta-) modeling framework, its components and their interaction, (3) evaluation and verification of the framework and the defined requirements.\\
\textbf{Meta-Language:} (1) Definition of the meta-language. \\
\textbf{Qualifiable Runtime:} (1) Selection of a suitable programming and verification language, (2) definition and implementation of the interface for interaction with models on M2 and M1 level, (3) definition and implementation of deterministic meta-model changes (reflections), (4) implementation of a demonstrator. \\
\textbf{Persistence: } (1) Defining an appropriate data-structure, (2) implementation of a suitable serialization.

\begin{figure}[!t]
    \centering
    \includevisio[width=2.7in]{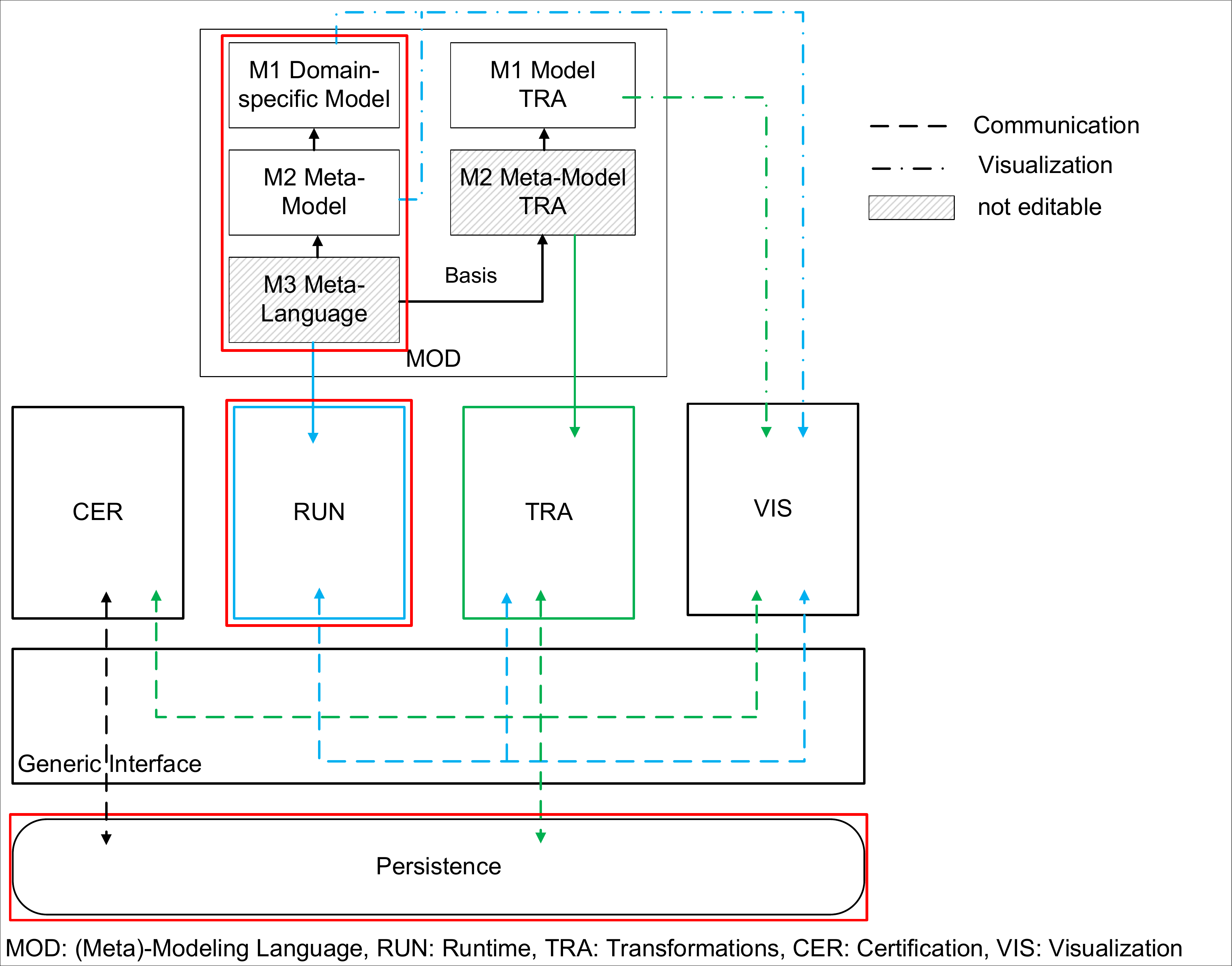}
    \caption{Interaction between components of (meta)-modeling framework \cite{METACERT_VISION}}
    \label{fig:interface}
\end{figure}

\textbf{Whole Framework:} As a first step towards a (meta-) modeling framework the interaction of all components of the framework was defined as it is illustrated in Figure \ref{fig:interface}. Dashed lines represent the communication between the different components over a generic interface, dotted-dashed lines indicate the visualization of the different models. Shaded components are not editable and colors are used to distinguish the different lines.
 The whole framework is as much as possible decoupled and modular. Due to the modularity, the components can be qualified independently from each other, or if required simply replaced with other established solutions. With the exception of the modeling part (MOD), acting as a specification, communication between the components is conducted over the generic interface. The modeling part with the meta-language on M3 level and the transformation language on M2 level based on the meta-language must be implemented directly into source code to be able to create the corresponding M2 and M1 models or transformations at runtime, respectively. Even if the models located in RAM could be written directly to a persistence, the way to go is also via the generic interface. It is to be expected that the structure will not change at this level of detail. The parts outlined in red, are considered in this PhD-Thesis.
 
\begin{figure}[!th]
    \centering
    \includevisio[width=3in]{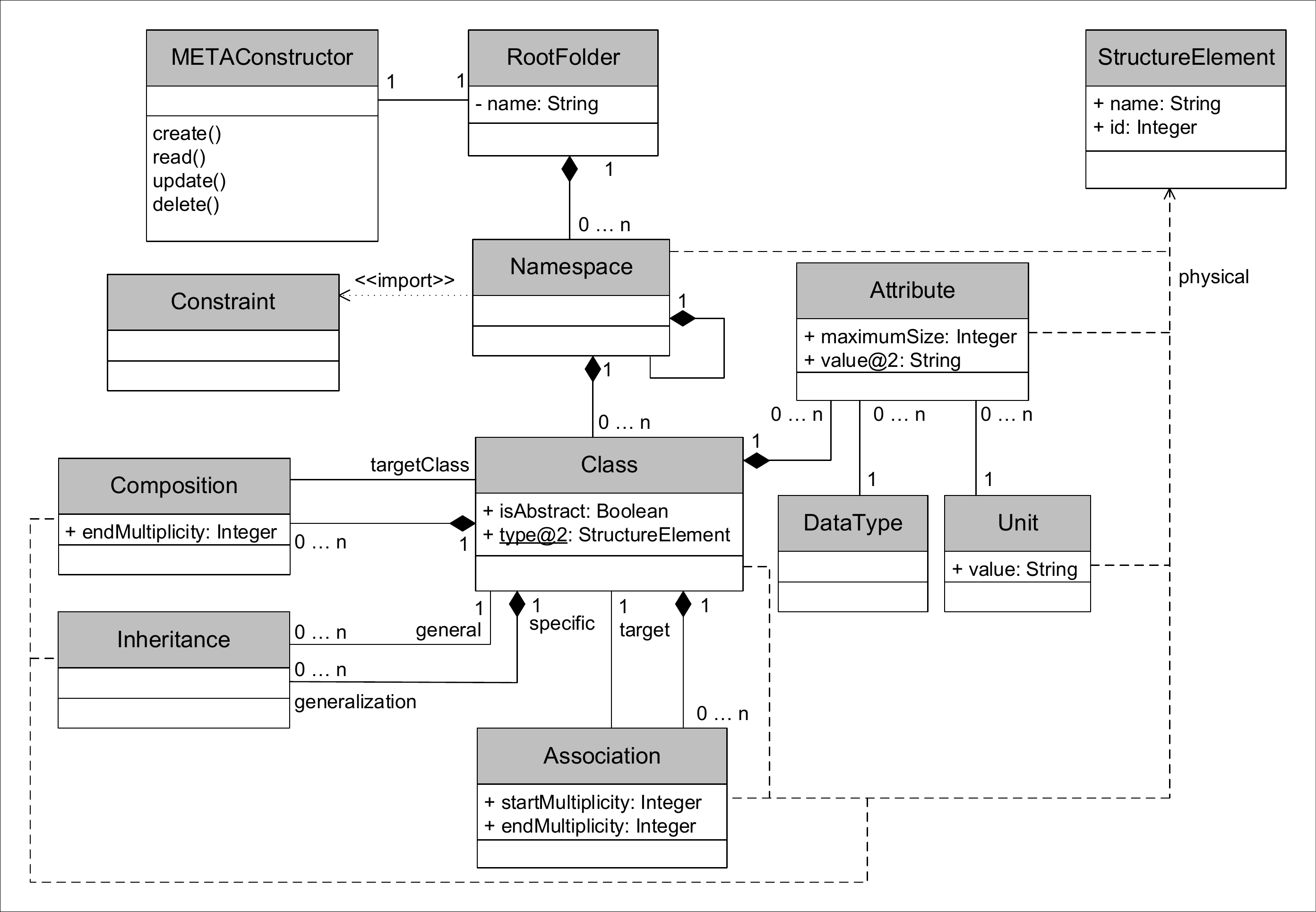}
    \caption{Meta-Language}
    \label{fig:meta_model}
\end{figure}

\textbf{Meta-Language:} A first prototype already exists for the meta-language which is depicted in Figure \ref{fig:meta_model}.
\newline
The main element is the \textit{Class} object in the middle of the UML chart. \textit{Classes} can be connected to each other via \textit{Compositions}, \textit{Inheritances} or \textit{Associations}. Unlike other languages, no separate base class is introduced for the references between classes, since these would have to be implemented using instance-of operators. The behavior of instance-of operators is not transparent and therefore increases the qualification effort of the implementation.  Every \textit{Class} can contain several \textit{Attributes} and every \textit{Attribute} has a \textit{DataType} and a \textit{Unit}. The \textit{Unit} is introduced to have an additional mechanism for checking transformations and to define the precise meaning of \textit{Attributes}. This could avoid errors that occur due to different understanding of the \textit{Attributes} or unit errors. For the ability of structuring and modularization meta-models, the \textit{Namespace} is introduced. The \textit{Namespace} can contain \textit{Classes} or other \textit{Namespaces}. The \textit{Constraint} element on the left of the \textit{Namespace} acts as a placeholder for a yet to be modeled constraint environment in order to define constraints on M2 level and to be able to check them on M1 level. The starting point of each model is the \textit{RootFolder}, in which the \textit{METAConstructor} operates in order to apply it to the complete model. The \textit{METAConstructor} interacts with the models on every level (M2 and M1) via CRUD-Requests. Even though typical UML elements are used, this is not intended to be based exclusively on the UML standard. In addition to the common UML notation, concepts of logical/ physical classification and deep-instantiation \cite{DeepInstantiation} as part of multi-level modeling are also utilized for the meta-language. This allows an unambiguous model-based specification of relations that are only assumed by convention in standard UML. Deep-instantiation enables the possibility of defining attributes at M3 level and instantiating them more then one level above at e.g. M1 level. In Figure \ref{fig:meta_model} it is depicted with the \textit{@} symbol and a number which indicates how many levels above the attribute can be instantiated. The \textit{@} symbol and the corresponding number are specified as potency value. The number is decremented on every modeling level above. Additionally, the \underline{underlined} attribute means that a value of this attribute has to be set on every level instead of only at the level where the potency value becomes zero. Without that concept it would only be possible to do instantiation one level above e.g. declaring on M3 and instantiating on M2 or declaring on M2 and instantiating on M1. Logical /physical classification is denoted with the \textit{physical} annotation, the dashed lines and the \textit{StructureElement} object in the UML chart. The \textit{StructureElement} is used to give every element on every modeling level the meaning of a physical object which always has a name and an identifier. Additionally, it enables the possibility of changing the type of each element during runtime because every element is at least a \textit{StructureElement} with a dedicated type defined a level below. The concepts described have very little impact on the users of the framework, e.g. it only forces the user to define the type of an object at both M2 and M1 level. 

\textbf{Qualifiable Runtime:} For the implementation of the meta-language, an appropriate programming language is required. Therefore, a comparison of the programming languages Ada SPARK, Ada, RUST and C was carried out. The focus was on a possible qualification of the programming language through concepts that facilitate the qualification process of the runtime e.g. prohibition of implicit type conversions, and the possibility of contract based programming. Due to its restrictivity and formal provability, Ada SPARK was utilized to implement the meta language. The result of formal proofs can support the qualification process. Determinism is primarily intended through the use of Ada SPARK as well as the avoidance of modeling constructs that cannot be implemented non-deterministically. The runtime acts only as a model interpreter in which the elements and the semantic of the meta language are implemented using Ada SPARK. The models (M2 and M1) are managed in RAM using an object-relational mapping approach. Each M3 element has its own fixed-size list of elements where the objects created at M2 level are stored. E.g. all objects of type class created at M2 level are stored in a list of all classes at runtime. Created attributes are again stored in a separate list and are assigned to the respective objects via global unique identifiers which do not change even when reloading the models. Each identifier is identical to the position in its corresponding list. This facilitates the handling of elements and their references in the runtime implementation.
For the interaction with the M2 and M1 models atomic CRUD-Requests are implemented as depicted in the following:
\begin{itemize}
    \item create $<$identifier$>$
    \item read $<$identifier$>$ $<$feature$>$
    \item update $<$identifier$>$ $<$feature$>$ $<$position$>$ $<$value$>$
    \item delete $<$identifier$>$.
\end{itemize}
Every request needs at least the identifier of the addressed element. The read request additionally needs the feature that has to be read e.g. name or type. For updating the value of a feature of an element the value and the position where the value has to be inserted is required. The position is needed to set values to certain positions in lists. If it is not a list, the position value 1 is selected. Currently, a connection to a higher order model query and modification language is being implemented. 

\textbf{Persistence:} So far, no progress has been made here.

\textbf{Proposed Planned Timeline and Future Work:}
The plan is to implement a first demonstrator with the programming language Ada SPARK in 2021. The prototype shall cover the whole functionality described within the current version of the meta-language including the creation of M2 level and M1 level models as well as the coupling to the higher order model query and modification language which is again coupled to an appropriate visualization. In addition, the first considerations proceed in the direction of storing the data in the persistence and implementing an appropriate serialization. Subsequently modeling and implementing a constraint environment within the meta-language for the creation and examination of constraints on M2 and M1 level in 2022. Furthermore, reflective requests shall be defined and implemented for interacting with and modifying the M2 level deterministically and in a way which ensures data integrity during runtime. 
In 2023, the focus should be mainly on the qualifiability of the runtime environment. To this end, requirements will be defined and reviewed. The requirements will be checked by means of defined scenarios and use cases. The elaboration of the thesis will take approximately one year and is scheduled for completion in 2024. 

\section*{Acknowledgment}
The German Federal Ministry for Economic Affairs and Energy (BMWi) has funded this research within the LUFO-VI program and the TALIA project. Additionally, I would like to thank my supervisor Bjoern Annighoefer for his guidance and support.

\newpage
\IEEEtriggeratref{20}
\bibliographystyle{IEEEtran}
\bibliography{IEEEabrv,doctoral_symposium}

\begin{thebibliography}{10}
\providecommand{\url}[1]{#1}
\csname url@samestyle\endcsname
\providecommand{\newblock}{\relax}
\providecommand{\bibinfo}[2]{#2}
\providecommand{\BIBentrySTDinterwordspacing}{\spaceskip=0pt\relax}
\providecommand{\BIBentryALTinterwordstretchfactor}{4}
\providecommand{\BIBentryALTinterwordspacing}{\spaceskip=\fontdimen2\font plus
\BIBentryALTinterwordstretchfactor\fontdimen3\font minus
  \fontdimen4\font\relax}
\providecommand{\BIBforeignlanguage}[2]{{%
\expandafter\ifx\csname l@#1\endcsname\relax
\typeout{** WARNING: IEEEtran.bst: No hyphenation pattern has been}%
\typeout{** loaded for the language `#1'. Using the pattern for}%
\typeout{** the default language instead.}%
\else
\language=\csname l@#1\endcsname
\fi
#2}}
\providecommand{\BIBdecl}{\relax}
\BIBdecl

\bibitem{Steinberg2009}
S.~Dave, B.~Frank, P.~Marcelo, and M.~Ed, \emph{EMF Eclipse Modeling
  Framework}.\hskip 1em plus 0.5em minus 0.4em\relax Addison-Wesley
  Professional, 2009.

\bibitem{GME}
A.~Ledeczi, M.~Maroti, A.~Bakay, G.~Karsai, J.~Garrett, C.~Thomason,
  G.~Nordstrom, J.~Sprinkle, and P.~Völgyesi, ``The generic modeling
  environment,'' \emph{Workshop on Intelligent Signal Processing, Budapest,
  Hungary}, vol.~17, 01 2001.

\bibitem{MATLAB_SIMULINK}
\BIBentryALTinterwordspacing
I.~The~MathWorks, ``Matlab simulink.'' [Online]. Available:
  \url{https://de.mathworks.com/products/simulink.html}
\BIBentrySTDinterwordspacing

\bibitem{MATLAB_SC}
\BIBentryALTinterwordspacing
------, ``Matlab system composer.'' [Online]. Available:
  \url{https://de.mathworks.com/products/system-composer.html}
\BIBentrySTDinterwordspacing

\bibitem{SCADE}
F.-X. Dormoy, ``Scade 6 a model based solution for safety critical software
  development,'' in \emph{Embedded Real Time Software and Systems (ERTS2008)},
  Jan. 2008.

\bibitem{DO178-C}
\emph{DO-178C Software Considerations in Airborne Systems and Equipment
  Certification}, {RTCA} Std., 2011.

\bibitem{ISO26262}
\emph{ISO 26262 Road Vehicles – Functional Safety}, {International
  Organization for Standardization} Std., 2018.

\bibitem{OOP_Issues}
S.~Subbiah and S.~Nagaraj, ``Issues with object orientation in verifying
  safety-critical systems,'' in \emph{Sixth {IEEE} International Symposium on
  Object-Oriented Real-Time Distributed Computing}.\hskip 1em plus 0.5em minus
  0.4em\relax {IEEE}, 2003.

\bibitem{DO332}
\emph{DO-332 Object-Oriented Technology and Related Techniques Supplement to
  DO-178C and DO-278A}, {RTCA} Std., 2011.

\bibitem{UML}
\emph{OMG Unified Modeling Language (OMG UML)}, {Object Management Group (OMG)}
  Std. 2.5.1, 2017.

\bibitem{MOF}
\emph{OMG Meta Object Facility (MOF) Core Specification}, {Object Management
  Group (OMG)} Std., 2019.

\bibitem{OAAM}
B.~Annighoefer, ``An open source domain-specific avionics system architecture
  model for the design phase and self-organizing avionics,'' in \emph{{SAE}
  Technical Paper Series}.\hskip 1em plus 0.5em minus 0.4em\relax {SAE}
  International, mar 2019.

\bibitem{SysML}
A.~Oliver, \emph{Modellbasierte Systementwicklung mit SysML}.\hskip 1em plus
  0.5em minus 0.4em\relax Carl Hanser Verlag GmbH Co KG, 2012.

\bibitem{AADL}
P.~H. Feiler, D.~P. Gluch, and J.~J. Hudak, ``The architecture analysis {\&}
  design language ({AADL}): An introduction,'' Defense Technical Information
  Center, Tech. Rep., feb 2006.

\bibitem{DO330}
\emph{DO-330 Software Tool Qualification Considerations}, {RTCA} Std., 2011.

\bibitem{DO331}
\emph{DO-331 Model-Based Development and Verification Supplement to DO-178C and
  DO-278A}, {RTCA} Std., 2011.

\bibitem{DO333}
\emph{DO-333 Formal Methods Supplement to DO-178C and DO-278A}, {RTCA} Std.,
  2011.

\bibitem{MISRAC}
\emph{MISRA-C:2004 - Guidelines for the use of the C language in critical
  systems}.\hskip 1em plus 0.5em minus 0.4em\relax MIRA, Limited, 2004.

\bibitem{TargetLink}
\BIBentryALTinterwordspacing
dSPACE GmbH, ``Targetlink,'' Dec. 2020. [Online]. Available:
  \url{www.dspace.com}
\BIBentrySTDinterwordspacing

\bibitem{Van_Acker_2020}
B.~V. Acker, B.~J. Oakes, M.~Moradi, P.~Demeulenaere, and J.~Denil, ``Validity
  frame concept as effort-cutting technique within theverification and
  validation of complex cyber-physical systems,'' \emph{Proceedings of the 23rd
  ACM/IEEE International Conference on Model Driven Engineering Languages and
  Systems: Companion Proceedings}, oct 2020.

\bibitem{VanAcker2018}
B.~Van~Acker, J.~Denil, and P.~De~Meulenaere, ``Generation of test strategies
  for model-based functional safety testing using an artifact-centric
  approach,'' 2018.

\bibitem{Zalila2013}
F.~Zalila, X.~Cr{\'{e}}gut, and M.~Pantel, ``Formal verification integration
  approach for {DSML},'' in \emph{Lecture Notes in Computer Science}.\hskip 1em
  plus 0.5em minus 0.4em\relax Springer Berlin Heidelberg, 2013, pp. 336--351.

\bibitem{Haxthausen2012}
A.~E. Haxthausen, ``Automated generation of safety requirements from railway
  interlocking tables,'' in \emph{Leveraging Applications of Formal Methods,
  Verification and Validation. Applications and Case Studies}.\hskip 1em plus
  0.5em minus 0.4em\relax Springer Berlin Heidelberg, 2012, pp. 261--275.

\bibitem{zalila2014methods}
F.~Zalila, ``Methods and tools for the integration of formal verification in
  domain-specific languages,'' Ph.D. dissertation, Institut National
  Polytechnique De Toulouse, 2014.

\bibitem{UPPAAL}
G.~Behrmann, A.~David, and K.~G. Larsen, ``A tutorial on uppaal,'' pp.
  200--236, 2004.

\bibitem{larsen1997uppaal}
K.~G. Larsen, P.~Pettersson, and W.~Yi, ``Uppaal in a nutshell,''
  \emph{International journal on software tools for technology transfer},
  vol.~1, no. 1-2, pp. 134--152, 1997.

\bibitem{OSATE}
\BIBentryALTinterwordspacing
``Osate.'' [Online]. Available: \url{https://osate.org/}
\BIBentrySTDinterwordspacing

\bibitem{Lasnier2009}
G.~Lasnier, B.~Zalila, L.~Pautet, and J.~Hugues, ``Ocarina : An environment for
  {AADL} models analysis and automatic code generation for high integrity
  applications,'' in \emph{Reliable Software Technologies {\textendash}
  Ada-Europe 2009}.\hskip 1em plus 0.5em minus 0.4em\relax Springer Berlin
  Heidelberg, 2009, pp. 237--250.

\bibitem{Besnard_2020}
V.~Besnard, F.~Jouault, M.~Brun, C.~Teodorov, P.~Dhaussy, and J.~Delatour,
  ``Modular deployment of uml modelsfor v$\&$v activities and embedded
  execution,'' oct 2020.

\bibitem{Besnard2018}
V.~Besnard, M.~Brun, F.~Jouault, C.~Teodorov, and P.~Dhaussy, ``Unified {LTL}
  verification and embedded execution of {UML} models,'' in \emph{Proceedings
  of the 21th {ACM}/{IEEE} International Conference on Model Driven Engineering
  Languages and Systems}.\hskip 1em plus 0.5em minus 0.4em\relax {ACM}, oct
  2018.

\bibitem{Hilken2014}
F.~Hilken, P.~Niemann, R.~Wille, and M.~Gogolla, ``Towards a base model for uml
  and ocl verification.'' in \emph{MoDeVVa@ MoDELS}.\hskip 1em plus 0.5em minus
  0.4em\relax Citeseer, 2014, pp. 59--68.

\bibitem{METACERT_VISION}
V.~Tietz, J.~Schöpf, A.~Waldvogel, and B.~Annighöfer, ``A concept for a
  qualifiable (meta)-modeling framework deployable in systems and tools of
  safety-critical and cyber-physical environments,'' in-press.

\bibitem{DeepInstantiation}
\BIBentryALTinterwordspacing
C.~Atkinson and T.~K\"{u}hne, ``Rearchitecting the {UML} infrastructure,''
  \emph{ACM Trans. Model. Comput. Simul.}, vol.~12, no.~4, pp. 290--321, Oct.
  2002. [Online]. Available: \url{https://doi.org/10.1145/643120.643123}
\BIBentrySTDinterwordspacing

\end{thebibliography}

\end{document}